# Spectroscopic evidence for electron correlations in the interface-modulated epitaxial bilayer graphene

Chaofei Liu[1] & Jian Wang[1,2,3,*]

[1] International Center for Quantum Materials, School of Physics, Peking University, Beijing 100871, China
[2] CAS Center for Excellence in Topological Quantum Computation, University of Chinese Academy of Sciences, Beijing 100190, China
[3] Beijing Academy of Quantum Information Sciences, Beijing 100193, China

[*]E-mail: jianwangphysics@pku.edu.cn

**Abstract**

Superlattice potentials are theoretically predicted to modify the single-particle electronic structures. The resulting Coulomb-interaction-dominated low-energy physics would generate highly novel many-body phenomena. Here, by *in situ* tunneling spectroscopy, we show the signatures of superstructure-modulated correlated electron states in epitaxial bilayer graphene (BLG) on 6*H*-SiC(0001). As the carrier density is locally quasi-'tuned' by the superlattice potentials of a 6×6 interface reconstruction phase, the spectral-weight transfer occurs between the two broad peaks flanking the charge-neutral point. Such detected non-rigid band shift beyond the single-particle band description implies the existence of correlation effects, probably attributed to the modified interlayer coupling in epitaxial BLG by the 6×6 reconstruction as in magic-angle BLG by the Moiré potentials. Quantitative analysis suggests the intrinsic interface reconstruction shows a high carrier tunability of ~½ filling range, equivalent to the back gating by a voltage of ~70 V in a typical gated BLG/SiO$_2$/Si device. The finding in interface-modulated epitaxial BLG with reconstruction phase extends the BLG platform with electron correlations beyond the magic-angle situation, and may stimulate further investigations on correlated states in graphene systems and other van der Waals materials.

## 1. Introduction

For twisted heterostructures between atomically thin van der Waals crystals, the electronic structures are modified by the Moiré superlattice potentials via interlayer hybridization. In the two-dimensional (2D) mini Brillouin zone for Moiré lattice, the continuum model of twisted bilayer graphene (BLG) predicted two low-energy bands separated by a gap from the higher-energy dispersive bands [1]. Near the 'magic angle' (MA) $\theta \approx 1.05°$, the effective non-Abelian gauge potentials generated by interlayer hoppings dictate the low-energy bands be highly non-dispersive (flat bands) near the charge-neutral point [1-3]. In such limit of strong electron correlations, the Coulomb interaction localizes the electrons, yielding emergent many-body ground states with high gate tunability near the fractional fillings of Moiré band, $v = 0, \pm¼, \pm½, \pm¾$, including Mott-correlated insulating states [4-7], superconductivity [5,6,8], ferromagnetism [6,9], and quantum anomalous Hall effect [10]. Particularly, at half-filling, the Mott-correlated phases and derived superconductivity upon slight carrier doping show a transport phase diagram resembling the high-$T_c$ cuprate superconductors [4-6].

Despite the intensive literature already reported, many fundamental questions desire further explorations. i) By integrating STM with the back-gate tuning of carrier doping, the signatures of electronic correlations in MA-TBG were detected behaving as distorted tunneling spectra (d$I$/d$V$ vs. $V$) upon the alignment of flat band with Fermi level [11,12]. Developing a different *in situ* carrier-tuning method would be beneficial to the spectroscopic investigations of correlated physics. ii) Signatures of correlated phase have been observed by both transport and scanning tunneling spectroscopy (STS) in BLG beyond MA of ~1.05° (e.g. $\theta = 0.7°$ [13]; $\theta = 0.93°$ [14]; $\theta = 1.49°$ [15]). In principle, twist angle and interlayer hybridization equivalently control the electronic structures of Moiré crystals [5]. By tuning interlayer spacing with hydrostatic pressure (1.33 GPa), non-MA-BLG ($\theta = 1.27°$) can also show the strong insulating phase, superconductivity, and fine structures of Landau fans as unpressured MA-BLG ($\theta = 1.08°$) [5]. It would be interesting to further explore the correlated states in BLG without MA.

Previously, the signature of correlated phase has been detected in epitaxial pristine monolayer graphene [16,17], which appears as the considerable deviation from the noninteracting Dirac-cone dispersion in photoemission-measured band structures. However, the electronic many-body correlations have not been reported in epitaxial bilayer graphene. Here, by *in situ* tuning the local carrier density alternatively via intrinsic interface reconstruction, we detected for the first time the STS signatures of correlated electron phase in epitaxial BLG on 6*H*-SiC(0001). Specifically, the 6×6 interfacial reconstruction between BLG and SiC was found to effectively modulate the spatial carrier density locally. As the Fermi energy is 'tuned' quasi-continuously, the spectral-weight transfer occurs across the Dirac point. Such observation deviating from the rigid band shift is beyond the noninteracting description and suggests the existence of electron correlations, which highlights the important role of interfacial 6×6 reconstruction in modifying the electronic structures of BLG.

## 2. Results

### 2.1. Epitaxially grown BLG on SiC

All STM experiments were performed at 4.3 K unless specified. The BLG was synthesized via surface graphitization of *n*-type 6*H*-SiC(0001) by flash annealing at 1300 ℃ for 80 cycles [18]. Typically, the graphitized SiC shows step heights of 0.75 nm [Fig. 1(a)]. Essentially, the triple bilayer-SiC (0.25×3 nm) steps uniformly capped by BLG naturally result in the 0.75-nm-high steps [18,19]. The primarily obtained epitaxial BLG is Bernal-stacked [Fig. 1(d)] with a triangular atomic lattice [Fig. 1(b)]. Fast-Fourier-transformation (FFT) analysis [upper inset of Fig. 1(b)] yields the lattice constant $a_0 \approx 0.26$ nm, comparable with $a_0 = 0.246$ nm for pristine graphene.

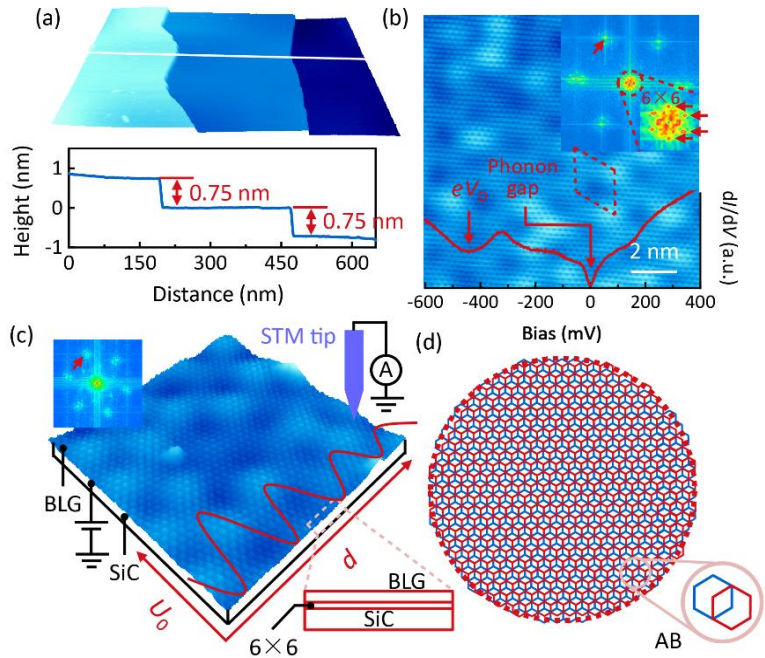

FIG. 1. Epitaxial BLG on 6*H*-SiC(0001). (a) Upper: large-scale topographic image of graphitized SiC (size: 650×650 nm$^2$; set point: $V = 1$ V, $I = 500$ pA); lower: linecut profile along the white line in upper panel. (b) Topographic image of the epitaxial BLG on SiC (size: 12×12 nm$^2$; set point: $V = 0.4$ V, $I = 500$ pA), showing the triangular atomic lattice accompanied by 6×6 reconstruction phase (supercell indicated as dashed diamond). Upper inset: FFT of the main panel, where the 6×6 features are zoomed in and highlighted by horizontal arrows. Lower inset: typical tunneling spectrum; the 'phonon' gap (see main text) at $E_F$ and the Dirac point $eV_D$ at local differential-conductance minimum are highlighted [set point: $V = 0.04$ V, $I = 500$ pA; modulation: $V_{mod} = 6$ mV (by default unless specified)]. (c) Schematic of the STM setup, integrated with the small-scale topography of BLG (size: 5×5 nm$^2$; set point: $V = 0.4$ V, $I = 500$ pA). Upper inset: FFT of the main panel. Lower inset: schematic profile of BLG/SiC with the interface 6×6 reconstruction. $U_0$, electrostatic potential; $d$, distance. (d) Bernal (AB) stacking of the BLG on SiC.

Tunneling spectra were acquired by the standard lock-in technique [20] [for schematic of STM apparatus, see Fig.

1(c)]. Distinctive gap- and dip-like anomalies are revealed at Fermi energy $E_F$ and [−500, −400] mV, respectively [lower inset of Fig. 1(b)]. The gap is attributed to the absence of phonon-mediated inelastic tunneling within ±Ω (Ω, phonon threshold energy) [21], whereas the dip that appears as a local differential-conductance minimum is indicative of the Dirac-type charge-neutral point nominally at $eV_D$ [21]. The Fermi level shifted 400–500 meV above $eV_D$ is well comparable to previously reported epitaxial graphene/SiC [16,22,23], suggesting the highly electron-doped nature of the epitaxial BLG. In essence, the $E_F$ shift is attributed to the doping of graphene layer by the depletion of surface electrons at the buffer layer–n-type SiC interface [24]. The interlayer coupling in Bernal-stacked BLG breaks the A/B sublattice symmetry of individual graphene layers. The resulting electronic structures are gapped with parabolic dispersions, yielding massive chiral Dirac fermions [25]. This contrasts the linear dispersion of massless Dirac carriers in monolayer graphene, and explains the nonlinear local density of states (LDOS) in the BLG spectrum near the charge-neutral Dirac point [lower inset of Fig. 1(b)].

*2.2. Interface-modulated spatial electron density*

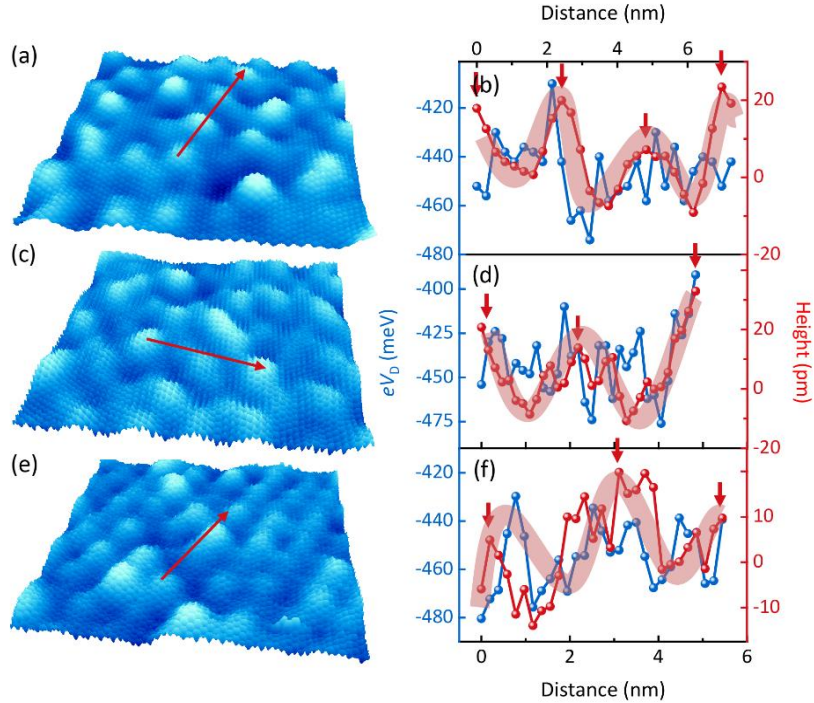

FIG. 2. Nominal Dirac-point energy modulated by the 6×6 reconstruction phase. (a,c,e) Topographic images of BLG in different positions. Size: (a,c,e), 12.5×12.5 nm$^2$; set point: a, $V$ = 0.3 V, $I$ = 500 pA, b, $V$ = 0.35 V, $I$ = 500 pA, c, $V$ = 0.4 V, $I$ = 500 pA. (b,d,f) Spatial dependences of $eV_D$ and height along the arrows in (a,c,e), respectively. The short arrows mark the approximate positions of the local height maxima. The shadowed wide strips are guides to the eyes of variations of $eV_D$ and height.

Note that a reconstruction pattern exists in BLG periodically [Fig. 1(b)], which follows the morphology of the C-rich 6×6 interface reconstruction beneath BLG with respect to the surface unit cell of SiC(0001) ($a_0 \approx 0.307$ nm) [lower inset of Fig. 1(c)] [26,27]. The existence of 6×6 reconstruction pattern is further supported by the corresponding low-momentum hexagonal spots in the FFT image of BLG topography [marked by '6×6', upper inset of Fig. 1(b)]. In the flash-annealing procedure for SiC graphitization, the 6×6 superstructure develops at the initial stage of graphene formation, serving as the buffer layer for graphene epitaxy. Compared with graphene, the 6×6 reconstruction buffer layer has the same honeycomb-type network of $sp^2$-derived σ bands, but without the graphene-like π bands responsible for the relativistic Dirac fermions [28]. Theoretically, the local topography curvature can induce spatially varying electrochemical potential [29]. In addition, in a twisted Moiré superlattice, the LDOS is modulated in space with the same period as the Moiré pattern [30]. Following these spirits, despite the negligible interaction between BLG and SiC [31], the 6×6 interface reconstruction phase would similarly generate effective periodic potential field in BLG and locally modulate the LDOS. Given that the 6×6 reconstruction arises from the periodic C–Si bonding strength [32], the spatially modulated LDOS is essentially an intrinsic effect for the graphene epitaxially grown on SiC.

In STM configurations, $I(V) \propto e^{-\frac{2}{\hbar}\sqrt{2m\phi}\Delta d} \int_{-eV}^{0} d\varepsilon \rho_s(\varepsilon + eV)$ ($\phi$, tunneling potential barrier; $\Delta d$, tip–sample distance; $\rho_s$, sample LDOS), meaning that the STM topographic image carries the information of integrated sample LDOS, $\int_{-eV}^{0} d\varepsilon \rho_s(\varepsilon + eV)$, besides measuring the topographic fluctuations $\Delta d(x,y)$. From the $I(V)$ formula, we can see that, for atomically flat sample surface, which is the situation of our epitaxial BLG, the integrated sample LDOS reflected in the measured STM topographic image is nonnegligible, because the linear dependence of $I(V)$ on integrated sample LDOS prevails over the exponential dependence of $I(V)$ on $\Delta d$ when $\Delta d(x,y)$ remains approximately constant. Based on these arguments, as in the Moiré superlattice, the observed periodic STM topographic fluctuations in BLG induced by 6×6 reconstruction phase is mainly attributed to the potential-field-modulated LDOS of BLG [Fig. 1(c)]. Accordingly, the LDOS would locally modulate the electron density $n$, and thus, effectively modulate the Fermi level $E_F$ and the nominal Dirac-point energy $eV_D$. Consistently, in our experiments, along different trajectories across the 6×6-reconstructed topographic 'humps', $eV_D$ and the LDOS-dominated 'height' *roughly* change synchronously (Fig. 2). For quantitative analysis, the data points of $eV_D$ are plotted against height, showing a statistical trend towards their positive correlation overall (Fig. S1). These 'entangled' $eV_D$ and height profiles establish the pronounced charge-density modulation intimately controlled by the 6×6 interface reconstruction.

*2.3. Electron correlations revealed by spectral-weight transfer*

The interface-reconstructed, non-gated BLG behaves equivalently as a periodic array of spatially doping-evolving back-gated 'mini-devices'. The charge-neutral point $eV_D$ extracted from STS has been commonly used as a local measure of the charge density because of its charge sensitivity [21]. Preliminarily, for typical tunneling spectra with different $eV_D$ (i.e. different dopings), we found the spectral intensities flanking $eV_D$ appear mutually complementary [Fig. 3(a)]. More specifically, as $eV_D$ increases approaching $E_F$ [light → dark red curve in Fig. 3(a)], the intensity of the lower band (LB) below $eV_D$ roughly increases. Meanwhile, the intensity of the upper band (UB) right above $eV_D$ decreases accordingly. Since the 6×6-superlattice-potential-tuned LDOS can modulate $eV_D$ via modulating the electron density $n$, as shown in Fig. 3(b), by ordering the locally measured tunneling spectra by $eV_D$, the electron density can be regarded as being effectively 'tuned' quasi-continuously. On the whole, the complementary effect can be also seen for the spectral intensities flanking $eV_D$.

Purely according to the false-color plot [Fig. 3(b)], the evolutionary trend of LB and UB appears relatively weak. To quantify the spectral intensities of LB and UB for quantitatively describing the complementary effect, the tunneling spectra were tentatively fitted by a multi-Gaussian function (Fig. S2). The spectral weights of the Gaussian components for LB and UB were extracted separately by integral as $w_{LB}$ and $w_{UB}$ [Fig. 3(c)]. Figure 3(d) plots $w_{LB}$ and $w_{UB}$ as functions of $eV_D$. Evidently, $w_{LB}$ and $w_{UB}$ show positive and negative correlations with $eV_D$, respectively, consistent with the qualitative judgment based on Fig. 3(b). These phenomena are highly reproducible for several different sets of spatially resolved tunneling spectra (Fig. S3). The contrast of $eV_D$ dependences for $w_{LB}$ and $w_{UB}$ directly suggests the local-charge-variation-tuned spectral-weight transfer between LB and UB [Fig. 3(e)].

Previously, in a back-gated MA-BLG ($\theta = 1.07°$) device, the spectral-weight transfer induced by gate tuning was attributed to a signature of Mott-type correlations [33]. In pristine Bernal-stacked BLG, signatures of spectral-weight transfer across the charge-neutral point have similarly shown up by the macroscopic carrier tuning via gate voltage [34], although further analysis is absent. In the weak-coupling mean-field picture, tuning the Fermi energy via carrier doping purely results in the rigid band shift in the single-particle electronic structures, where the kinetic energy $U_t$ at Fermi energy $E_F$ is larger than the Coulomb interaction $U$. The detected spectral-weight transfer is beyond the noninteracting, single-particle description, and essentially can be attributed to the electron-correlation effects. Note that while the neutrality point $eV_D$ increases as local carrier tunning, UB and LB do not shift in energy following $eV_D$. Such non-rigid band-shift behavior beyond single-particle scenario can be consistently incorporated within the correlation explanation.

A thorough theoretical description of the observations remains lacking. In phenomenology, the interface-modulated spectral-weight transfer here is reminiscent of the doping-induced spectral-weight redistribution between Hubbard bands and charge-transfer gap in a doped Mott insulator [35,36]. Such resemblance implies the Mott-like correlated phase as a candidate explanation (Part SI). More concretely, the detected electron-correlation effect can be induced by the plasmarons derived from the electron–plasmon coupling for the electronic states near Dirac point, as previously established in pristine quasi-free-standing graphene on SiC by photoemission spectroscopy [Fig. 3(e); Part SII] [16,17].

*2.4. High carrier tunability of 6×6 interfacial reconstruction phase*

For the back-gated MA-BLG ($\theta = 1.07°$), a gate voltage of $\Delta V_g \approx \pm 15$ V shows a carrier-tuning ability of half-filling $\Delta \nu = \pm \frac{1}{2} n_s$ ($n_s = 4/A_0 = 2.7 \times 10^{12}$ cm$^{-2}$, i.e. four carriers per Moiré supercell $A_0$) of the fourfold spin–valley degenerate flat band [33]. Note that the degree of spectral-weight transfer therein provides a sensitive measure of the local charge. Semi-quantitatively, the local filling fraction $\nu$ can be estimated from the spectral weights of LB and UB, i.e., $\nu = 2\left(\frac{w_{LB}}{w_{LB}+w_{UB}} - \frac{1}{2}\right)$ [33]. For epitaxial BLG, $\nu$ ('pseudo-filling') within LB below $E_F$ was similarly defined purely for quantifying the carrier tunability of 6×6 reconstruction. Figure 3(f) presents the extracted $\nu$ plotted vs. $eV_D$. Profoundly, the intrinsically available electron density modulated by the 6×6 reconstruction can access a filling range of $\nu = 1/8\, n'_s$–$5/8\, n'_s$ ($n'_s$, 'full filling' of LB). This highlights a striking carrier-tuning ability of the interface reconstruction highly comparable with the gating technique.

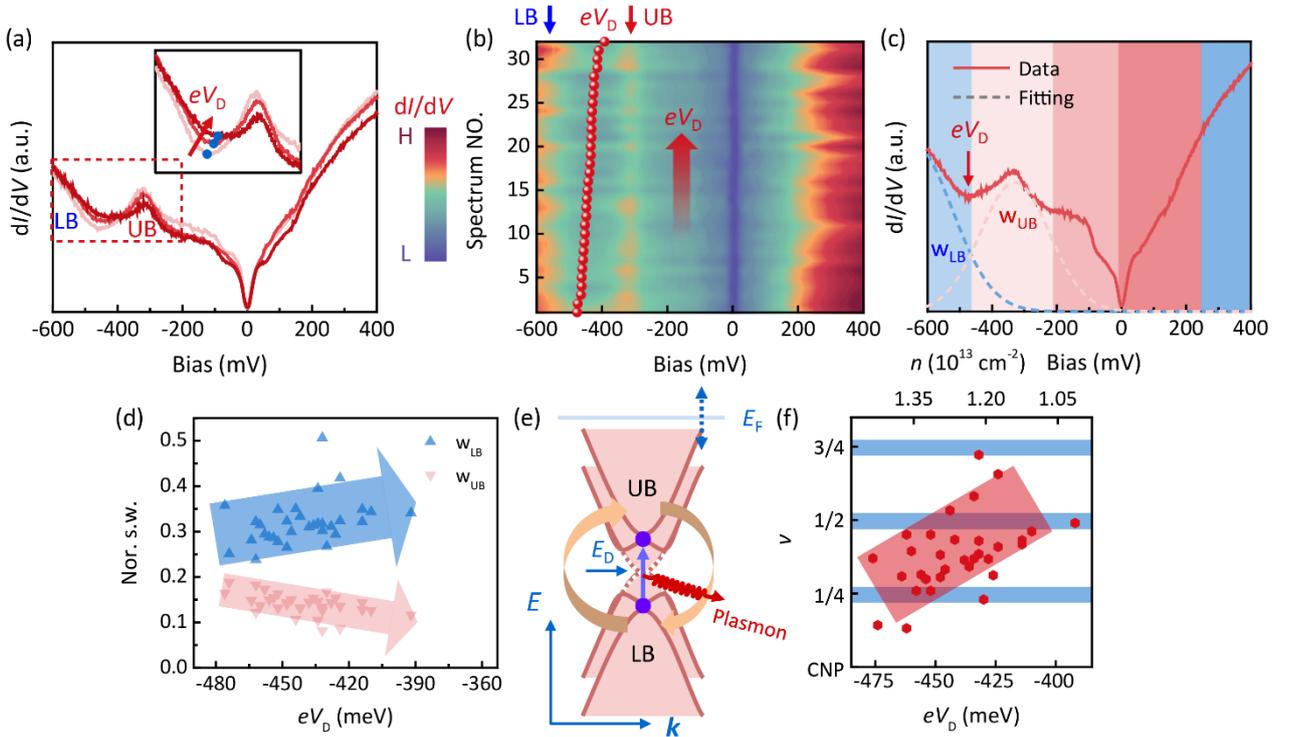

FIG. 3. Spectral-weight transfer in BLG. (a) Tunneling spectra at several typical $eV_D$ extracted from spectra collected at different positions. Inset: zoom-in view of the spectra within the dashed rectangle for more clearly showing the spectral evolution as $eV_D$ (blue dots) changes. The arrow highlights the $eV_D$-increasing direction among different spectra. (b) False-color plot of the tunneling spectra taken along the arrow in Fig. 2(c) ordered by increasing $eV_D$ as indicated by the arrow. (c) Gaussian components (dashed curves) for LB and UB flanking the Dirac point obtained from the multi-Gaussian fit to the tunneling spectrum (solid red curve). The integrated spectral weights are denoted as w$_{LB}$ and w$_{UB}$, respectively. (For all components involved in the multi-Gaussian fit, see Fig. S2.) (d) Normalized spectral weights (s.w.) w$_{LB}$ and w$_{UB}$ extracted from the spectra in (b) plotted vs. $eV_D$. (e) Schematic of the spectral-weight transfer between LB and UB, and plasmon emission for states near Dirac point. (f) $eV_D$ dependence of the filling factor $\nu$ within LB calculated from (d) as described In the main text.

By excluding the inelastic-tunneling effect involving the acoustic phonon $\Omega \approx 27$ meV (defined as half the 'phonon'-gap energy), the true Dirac-point energy $E_D$ in graphene band structures should be modified according to $|E_D| = e|V_D| - \Omega$ [21]. Further referring to the $n$–$E_D$ relation for ideal graphene, $n(r) = \frac{E_D^2(r)}{\pi(\hbar v_F)^2}$, we converted $eV_D$ to the local carrier density $n$ (assuming Fermi velocity $v_F = 10^6$ m/s) [Fig. 3(f)]. The estimated $n$ falls within $9.8 \times 10^{12}$–$1.5 \times 10^{13}$ cm$^{-2}$, corresponding to a strikingly high doping range of $\Delta n = 5.2 \times 10^{12}$ cm$^{-2}$ (= $1.9 n_s$ as in MA-BLG). In the gate-tunable BLG/SiO$_2$/Si device, based on a simple parallel-plate capacitor model, the carrier density $n$ directly scales with applied gate voltage $V_g$ as $\Delta n = \alpha \Delta V_g$. Here, $\alpha$ is determined by the gate capacitance, $\alpha = \frac{\varepsilon_0 \varepsilon}{te}$ ($\varepsilon_0, \varepsilon$, permittivities of free space and SiO$_2$, respectively; $t$, thickness of SiO$_2$), and concretely, $\alpha = 7.19 \times 10^{10}$ cm$^{-2}$ V$^{-1}$ for 300-nm SiO$_2$ [34]. Based on these discussions, the electron-density range $\Delta n$ for BLG/SiC here corresponds to $\Delta V_g = 72.3$ V, indeed comparable with the measured $\Delta V_g = 55$–85 V for tuning $\Delta n = 2n_s$ in BN/SiO$_2$-back-gated MA-BLG devices [11,33]. Therefore, the local charge modulation by 6×6 interface reconstruction can be physically equivalent to the back gating by a high voltage of ~70 V in a typical BLG/SiO$_2$/Si device.

## 3. Discussions

In atomically thin 2D van der Waals heterostructures, the interlayer-hybridized Moiré superlattice and the large effective mass of electrons are two crucial ingredients in the engineered Mott-correlated phase. For BLG/SiC, while the effective mass is unknown, the Moiré-equivalent interlayer hybridization by the 6×6 interface reconstruction in BLG remains predominately workable. Recall that in the hydrostatic-pressured non-MA-BLG ($\theta = 1.27°$) with tuned interlayer coupling, the correlated insulating state was observed 'unexpectedly' at half-filling, where superconductivity is induced by further hole doping [5]. The filling sequence of Landau level in this pressured non-MA-BLG is identical to that in Bernal-stacked BLG, where the low-energy bands are dominated by parabolic dispersions because of interlayer tunneling [25]. Low-energy bands touching quadratically at $K$ and $K$' points as Bernal BLG are indeed involved in the recent continuum model of MA-BLG flat bands [37]. From these perspectives, although the flat band is likely absent in Bernal BLG, the observation of electron correlations therein can be presumably expected, highlighting the importance of 6×6 interface superlattice potentials in modifying the interlayer hybridization. For decisive check over the role of the 6×6 reconstruction, control experiment may be recommended via intercalating H atoms between buffer layer and SiC to saturate the interface C–Si bonds for removing the 6×6 buffer phase [38].

The following difference may be noted between the detected electron-correlation effect in epitaxial BLG and MA-BLG. i) The electronic structures are modified by the interlayer hybridization tuned by the Moiré superlattice in MA-BLG, but tuned by the 6×6 superstructure in epitaxial BLG. ii) In MA-BLG, the electron correlations are revealed as the *whole* carrier density is continuously tuned by the gate voltage (e.g. into the fractional-filling region). Whereas in epitaxial BLG, the correlation effect is captured as the *local* carrier density is 'tuned' quasi-continuously by the 6×6 reconstruction. iii) The correlated electronic states attributed to the Mott physics normally occur near $E_F$ for MA-BLG. But the correlated electrons are observed near the Dirac point far below $E_F$ here, as the situation for the correlated electrons with plasmon emissions observed in the epitaxial monolayer graphene/SiC [16,17]. From this angle, the correlations revealed by weight transfer most probably arise from the electron–plasmon coupling.

Alternative explanation of the detected charge modulation and spectral-weight transfer may be the contributions from the local fluctuations of the in-built dipole field created across the BLG layers. During the sublimation process for preparing BLG, the interface between BLG and 6$H$-SiC(0001) crystal can become chemically inhomogeneous. Note that BLG epitaxially grown on SiC has a dipole field induced between the SiC depletion layer and the charge-accumulated graphene layer next to the interface. The two graphene layers in BLG are thus rendered inequivalent regarding the charge and the electrostatic potential. The chemically inhomogeneous interface would produce local changes in the surface dipole. This in turn would modulate the charge-neutrality Dirac point, and blur the statistical relation between 6×6 superstructure and $eV_D$ [Figs. 2(b), 2(d) and 2(f)] to a broadened scatter diagram (Fig. S1). Furthermore, because of the charging inequivalence of two layers in BLG, the UB and LB have different weights on the upper and lower layers. As a surface-sensitive technique, STM is more sensitive to the states of the outermost layer. Since the spectral weights of LB and UB depend on the dipole-field direction, the spectral-weight transfer can be expected from the local dipole-field fluctuations.

## 4. Perspectives

Our results by STM/STS suggest the previously overlooked many-body correlations are required to describe the low-energy properties of interface-modulated Bernal-stacked BLG. The observation extends the correlated BLG system beyond MA limitation, relaxing the requirement of precise angle controlling in twisted BLG. Different from the situation in Moiré superlattice with a twisted MA, where the correlated-insulator state and the superconductivity are mutually verified in transport $T$–$n$ diagram, the correlated phase in our experiments is visualized locally by the spectroscopic technique. Unconventional superconductivity arises from the Mott-correlated phase with slight carrier doping. Thus, it would be highly enlightening to the long-standing enigma of high-temperature superconductivity if the induced superconducting state can be directly further detected in space, e.g. by STM, in a parent correlated electronic system. Considering that the electrical transport involves the carriers within $k_BT$ of $E_F$, for sufficiently high charge density $[(E_F - E_D) > k_BT]$, the observed electron correlations will hardly affect the equilibrium transport properties. In the low-doping limit $[k_BT > (E_F - E_D)]$, carriers near the Fermi energy would be thermally excited and participate in transport. In consequence, the finding may stimulate explorations for additionally tracing the doping evolution towards the correlation-derived superconducting state at an atomic scale and establishing a more direct connection to high-$T_c$ cuprates beyond the purely phenomenological similarity in transport phase diagrams. In the future, the suppression of dipole fluctuations during growth process and the elaborate high-resolution STS data are highly desired to further clarify the interface reconstruction modulation and the corresponding electron correlations.

## Data availability statement

All data that support the findings of this study are included within the article (and any supplementary files).

## Acknowledgment

We thank Shan Zhong for the data analysis, Qingyan Wang, Ziqiao Wang, Cheng Chen, Yi Liu and Shusen Ye for the experimental assistance, and Allan H. MacDonald and Lin He for the fruitful discussions. This work was financially supported by National Natural Science Foundation of China (No.11888101, and No. 11774008), National Key R&D Program of China (No. 2018YFA0305604, and No. 2017YFA0303302), Beijing Natural Science Foundation (No. Z180010), and Strategic Priority Research Program of Chinese Academy of Sciences (No. XDB28000000).

## References

[1] Bistritzer R and MacDonald A H, 2011 Moiré bands in twisted double-layer graphene Proc. Natl. Acad. Sci. USA **108,** 12233.
[2] Trambly de Laissardiere G, Mayou D and Magaud L, 2010 Localization of dirac electrons in rotated graphene bilayers Nano Lett. **10,** 804.
[3] Suárez Morell E, Correa J D, Vargas P, Pacheco M and Barticevic Z, 2010 Flat bands in slightly twisted bilayer graphene: Tight-binding calculations Phys. Rev. B **82,** 121407.
[4] Cao Y, Fatemi V, Demir A, Fang S, Tomarken S L, Luo J Y, Sanchez-Yamagishi J D, Watanabe K, Taniguchi T, Kaxiras E, Ashoori R C and Jarillo-Herrero P, 2018 Correlated insulator behaviour at half-filling in magic-angle graphene superlattices Nature **556,** 80.
[5] Yankowitz M, Chen S, Polshyn H, Zhang Y, Watanabe K, Taniguchi T, Graf D, Young A F and Dean C R, 2019 Tuning superconductivity in twisted bilayer graphene Science **363,** 1059.
[6] Lu X, Stepanov P, Yang W, Xie M, Aamir M A, Das I, Urgell C, Watanabe K, Taniguchi T, Zhang G, Bachtold A, MacDonald A H and Efetov D K, 2019 Superconductors, orbital magnets and correlated states in magic-angle bilayer graphene Nature **574,** 653.
[7] Polshyn H, Yankowitz M, Chen S, Zhang Y, Watanabe K, Taniguchi T, Dean C R and Young A F, 2019 Large

linear-in-temperature resistivity in twisted bilayer graphene Nat. Phys. **15,** 1011.

[8] Cao Y, Fatemi V, Fang S, Watanabe K, Taniguchi T, Kaxiras E and Jarillo-Herrero P, 2018 Unconventional superconductivity in magic-angle graphene superlattices Nature **556,** 43.

[9] Sharpe A L, Fox E J, Barnard A W, Finney J, Watanabe K, Taniguchi T, Kastner M A and Goldhaber-Gordon D, 2019 Emergent ferromagnetism near three-quarters filling in twisted bilayer graphene Science **365,** 605.

[10] Serlin M, Tschirhart C, Polshyn H, Zhang Y, Zhu J, Watanabe K, Taniguchi T, Balents L and Young A, 2020 Intrinsic quantized anomalous Hall effect in a moiré heterostructure Science **367,** 900.

[11] Xie Y, Lian B, Jack B, Liu X, Chiu C L, Watanabe K, Taniguchi T, Bernevig B A and Yazdani A, 2019 Spectroscopic signatures of many-body correlations in magic-angle twisted bilayer graphene Nature **572,** 101.

[12] Choi Y, Kemmer J, Peng Y, Thomson A, Arora H, Polski R, Zhang Y, Ren H, Alicea J, Refael G, von Oppen F, Watanabe K, Taniguchi T and Nadj-Perge S, 2019 Electronic correlations in twisted bilayer graphene near the magic angle Nat. Phys. **15,** 1174.

[13] Kima K, DaSilvab A, Huangc S, Fallahazada B, Larentisa S, Taniguchid T, Kenji Watanabe B J L, MacDonaldb A H and Tutuca E, 2017 Tunable moiré bands and strong correlations in small-twist-angle bilayer graphene Proc. Natl. Acad. Sci. USA **114,** 3364.

[14] Codecido E, Wang Q, Koester R, Che S, Tian H, Lv R, Tran S, Watanabe K, Taniguchi T, Zhang F, Bockrath M and Lau C N, 2019 Correlated insulating and superconducting states in twisted bilayer graphene below the magic angle Sci. Adv. **5,** eaaw9770.

[15] Ren Y-N, Lu C, Zhang Y, Li S-Y, Liu Y-W, Yan C, Guo Z-H, Liu C-C, Yang F and He L, 2020 Spectroscopic Evidence for a Spin- and Valley-Polarized Metallic State in a Nonmagic-Angle Twisted Bilayer Graphene ACS nano **14,** 13081.

[16] Bostwick A, Ohta T, Seyller T, Horn K and Rotenberg E, 2007 Quasiparticle dynamics in graphene Nat. Phys. **3,** 36.

[17] Bostwick A, Speck F, Seyller T, Horn K, Polini M, Asgari R, MacDonald A H and Rotenberg E, 2010 Observation of plasmarons in quasi-freestanding doped graphene Science **328,** 999.

[18] Wang Q, Zhang W, Wang L, He K, Ma X and Xue Q, 2013 Large-scale uniform bilayer graphene prepared by vacuum graphitization of 6H-SiC(0001) substrates J. Phys. Condens. Matter **25,** 095002.

[19] Hupalo M, Conrad E H and Tringides M C, 2009 Growth mechanism for epitaxial graphene on vicinal 6*H*-SiC(0001) surfaces: A scanning tunneling microscopy study Phys. Rev. B **80,** 041401.

[20] Liu C, Wang G and Wang J, 2019 Manipulating the particle-hole symmetry of quasiparticle bound states in geometric-size-varying Fe clusters on one-unit-cell FeSe/SrTiO$_3$(001) J. Phys.: Condens. Matter **31,** 285002.

[21] Zhang Y, Brar V W, Wang F, Girit C, Yayon Y, Panlasigui M, Zettl A and Crommie M F, 2008 Giant phonon-induced conductance in scanning tunnelling spectroscopy of gate-tunable graphene Nat. Phys. **4,** 627.

[22] Brar V W, Zhang Y, Yayon Y, Ohta T, McChesney J L, Bostwick A, Rotenberg E, Horn K and Crommie M F, 2007 Scanning tunneling spectroscopy of inhomogeneous electronic structure in monolayer and bilayer graphene on SiC Appl. Phys. Lett. **91,** 122102.

[23] Zhou S Y, Gweon G H, Fedorov A V, First P N, de Heer W A, Lee D H, Guinea F, Castro Neto A H and Lanzara A, 2007 Substrate-induced bandgap opening in epitaxial graphene Nat. Mater. **6,** 770.

[24] Berger C, Song Z, Li T, Li X, Ogbazghi A Y, Feng R, Dai Z, Marchenkov A N, Conrad E H and First P N, 2004 Ultrathin epitaxial graphite: 2D electron gas properties and a route toward graphene-based nanoelectronics J. Phys. Chem. B **108,** 19912.

[25] McCann E and Fal'ko V I, 2006 Landau-level degeneracy and quantum Hall effect in a graphite bilayer Phys. Rev. Lett. **96,** 086805.

[26] Owman F and Mårtensson P, 1996 Scanning tunneling microscopy study of SiC(0001) surface reconstructions J. Vac. Sci. Technol. **14,** 933.

[27] Chen W, Xu H, Liu L, Gao X, Qi D, Peng G, Tan S C, Feng Y, Loh K P and Wee A T S, 2005 Atomic structure of the 6*H*–SiC(0001) nanomesh Surf. Sci. **596,** 176.

[28] Emtsev K V, Seyller T, Speck F, Ley L, Stojanov P, Riley J D and Leckey R C G, 2007 Initial Stages of the Graphite-SiC(0001) Interface Formation Studied by Photoelectron Spectroscopy Mater. Sci. Forum **556-557,** 525.

[29] Kim E-A and Neto A C, 2008 Graphene as an electronic membrane Europhys. Lett. **84,** 57007.

[30] Yankowitz M, Xue J, Cormode D, Sanchez-Yamagishi J D, Watanabe K, Taniguchi T, Jarillo-Herrero P, Jacquod P and LeRoy B J, 2012 Emergence of superlattice Dirac points in graphene on hexagonal boron nitride Nat. Phys. **8,** 382.

[31] Seyller T, Emtsev K V, Speck F, Gao K Y and Ley L, 2006 Schottky barrier between 6H-SiC and graphite: Implications for metal/SiC contact formation Appl. Phys. Lett. **88,** 242103.

[32] Kim S, Ihm J, Choi H J and Son Y W, 2008 Origin of anomalous electronic structures of epitaxial graphene on silicon carbide Phys. Rev. Lett. **100,** 176802.

[33] Jiang Y, Lai X, Watanabe K, Taniguchi T, Haule K, Mao J and Andrei E Y, 2019 Charge order and broken rotational symmetry in magic-angle twisted bilayer graphene Nature **573,** 91.

[34] Rutter G M, Jung S, Klimov N N, Newell D B, Zhitenev N B and Stroscio J A, 2011 Microscopic polarization in bilayer graphene Nat. Phys. **7,** 649.

[35] Eskes H, Meinders M B and Sawatzky G A, 1991 Anomalous transfer of spectral weight in doped strongly correlated systems Phys. Rev. Lett. **67,** 1035.

[36] Cai P, Ruan W, Peng Y, Ye C, Li X, Hao Z, Zhou X, Lee D-H and Wang Y, 2016 Visualizing the evolution from the Mott insulator to a charge-ordered insulator in lightly doped cuprates Nat. Phys. **12,** 1047.

[37] Hejazi K, Liu C, Shapourian H, Chen X and Balents L, 2019 Multiple topological transitions in twisted bilayer graphene near the first magic angle Phys. Rev. B **99,** 035111.

[38] Riedl C, Coletti C, Iwasaki T, Zakharov A A and Starke U, 2009 Quasi-free-standing epitaxial graphene on SiC obtained by hydrogen intercalation Phys. Rev. Lett. **103,** 246804.

# Spectroscopic evidence for electron correlations in the interface-modulated epitaxial bilayer graphene

Chaofei Liu[1] & Jian Wang[1,2,3,*]

[1] International Center for Quantum Materials, School of Physics, Peking University, Beijing 100871, China
[2] CAS Center for Excellence in Topological Quantum Computation, University of Chinese Academy of Sciences, Beijing 100190, China
[3] Beijing Academy of Quantum Information Sciences, Beijing 100193, China

[*] E-mail: jianwangphysics@pku.edu.cn

**SUPPLEMENTARY TEXT**

**I. Discussion About Mott Correlated Phase as the Candidate Origin of Detected Spectral-Weight Transfer**

In the Mott–Hubbard model for $d$-electron systems, the interplay of on-site Coulomb energy $U$, kinetic energy $U_t$, bandwidth $W$, and filling factor $v$ cooperatively drives the Mott-insulator transition [1]. The strong electron correlations are set by the high ratio of the on-site Coulomb interaction $U$ to the bandwidth $W$ of the flat band. In the Mott–Hubbard scenario, the strong on-site Coulomb repulsion forbids neighboring electron hopping; further at half-filling, Mott insulator occurs as the ground state with a correlated charge gap between the 'splitted' upper and lower Hubbard bands derived from the $d$-electron band [1].

Experimentally, in the doped Mott phase of cuprates, as increasing the doping level, the spectral weight transfers from Hubbard bands to charge-transfer gap near $E_F$ as in-gap excitations [2,3]. These excitation modes 'straddle' the simultaneously formed sharp V-shaped gap at $E_F$. With further carrier doping towards the underdoped superconducting (SC) regime, the V-shaped gap can finally evolve into the SC gap [3]. The resemblance between our result and such spectral-weight transfer in a doped Mott insulator indicates the Mott correlation as a likely explanation.

In the candidate Mott picture for our BLG, the Dirac-type conductance dip, where the charge-neutral point is located and where the LB/UB overlap, should be a correlated gap separating the Hubbard bands by $U$ in essence. Physically, the relative spectral weights of LB and UB flanking the correlated gap sensitively depend on the local carrier density, naturally reconciling the observed spectral-weight transfer behavior [Fig. 3(d)].

**II. Electron Correlations Arising from Electron–Plasmon Coupling**

We now discuss the possible concrete mechanism responsible for the detected electron correlations. Theoretically, plasmarons have been proposed in 2D electron gas defined as the resonantly coupled electron–plasmon composite excitations [4]. In the spectral function $A(k,E)$ and the LDOS spectra, the signal of the plasmaron appears as a satellite peak aside from the most tightly bounded conventional quasiparticle peak (main peak) [4,5]. By increasing chemical doping and/or disorder, the plamaron signature is expected to be suppressed [4,5].

For graphene, previous study shows that the carrier decaying via plasmon scattering is a well-defined quasiparticle only outside the region of kinematically allowed electron–hole excitations in $(E,k)$ phase diagram [6]. This constraint causes that the states near Dirac point interact strongly with the plasmons [Fig. 3(d)], meaning that the plasmarons can have a large influence on the many-body effect around $E \sim E_D$ [6,7]. Consistently, in experiments by angle-resolved photoemission spectroscopy (ARPES), the lower-energy plasmaron satellite band and the higher-energy main band straddling the Dirac point far below $E_F$ (~−0.5 eV) have been detected in the spectral function of quasi-free-standing graphene, where the plasmaron band appears suppressed by higher chemical doping [8]. Furthermore, while the interactions between quasiparticles and plasmons are stronger in the 2D massless Dirac system than in an ordinary 2D parabolic-band system [7], the plasmaron in a 2D massive-electron system (e.g. GaAs quantum well) is still observed by tunneling spectroscopy [9]. All these arguments suggest that, the LB near Dirac energy in our BLG, which behaves as the satellite band of UB and is suppressed as increasing electron doping (more negative $eV_D$) [Figs. 3(a)–3(c)], can be interpreted as the plamaron band. Equivalently speaking, the electron correlations probably arise because of the plasmon emissions for electrons near the Dirac point [Fig. 3(d)], i.e., the plasmarons consisting of resonantly bounded electrons and plasmons.

Compared with the plasmaron peak in quasi-free-standing graphene directly grown epitaxially on SiC [8], the plasmaron band in ordinary epitaxial graphene on buffer layer/SiC is normally not well resolvable in the ARPES-measured spectral function due to the high dielectric screening (with effective dielectric constant ε 3–4 times higher), which is likely further damped by the disorder in the sample. Thus, the spectral-weight transfer indicated by STS provides an alternative method to effectively reveal the many-body effect induced by the electron–plasmon coupling in ordinary epitaxial graphene.

Note that the plasmon emission occurs for the states near $E_D$ [6]. The gap opening at $E_D$ for epitaxial BLG would weaken the plasmaron response. Furthermore, the plasmaron band usually occurs at an energy roughly proportional to $E_F$ or carrier density [8,10]. Yet, the energies of LB and UB here appear approximately unchanged with doping as shifting $E_F$ (such absence of rigid following for LB and UB is also an indication of the existence of electron correlations) [Fig. 3(a)], which is left for future explorations for a full understanding within the plasmaron scenario.

**REFERENCES**
[1] Imada M, Fujimori A and Tokura Y, 1998 Metal-insulator transitions Rev. Mod. Phys. **70,** 1039.
[2] Eskes H, Meinders M B and Sawatzky G A, 1991 Anomalous transfer of spectral weight in doped strongly correlated systems Phys. Rev. Lett. **67,** 1035.
[3] Cai P, Ruan W, Peng Y, Ye C, Li X, Hao Z, Zhou X, Lee D-H and Wang Y, 2016 Visualizing the evolution from the Mott insulator to a charge-ordered insulator in lightly doped cuprates Nat. Phys. **12,** 1047.
[4] Jalabert R and Das Sarma S, 1989 Many-polaron interaction effects in two dimensions Phys. Rev. B **39,** 5542.
[5] von Allmen P, 1992 Plasmaron excitation and band renormalization in a two-dimensional electron gas Phys. Rev. B **46,** 13345.
[6] Bostwick A, Ohta T, Seyller T, Horn K and Rotenberg E, 2007 Quasiparticle dynamics in graphene Nat. Phys. **3,** 36.
[7] Polini M, Asgari R, Borghi G, Barlas Y, Pereg-Barnea T and MacDonald A H, 2008 Plasmons and the spectral function of graphene Phys. Rev. B **77,** 081411.
[8] Bostwick A, Speck F, Seyller T, Horn K, Polini M, Asgari R, MacDonald A H and Rotenberg E, 2010 Observation of plasmarons in quasi-freestanding doped graphene Science **328,** 999.
[9] Dial O, Ashoori R, Pfeiffer L and West K, 2012 Observations of plasmarons in a two-dimensional system: Tunneling measurements using time-domain capacitance spectroscopy Phys. Rev. B **85,** 081306.
[10] Dial O E, 2007 Single Particle Spectrum of the Two Dimensional Electron Gas thesis, Massachusetts Institute of Technology, Cambridge, MA.

**SUPPLEMENTARY FIGURES**

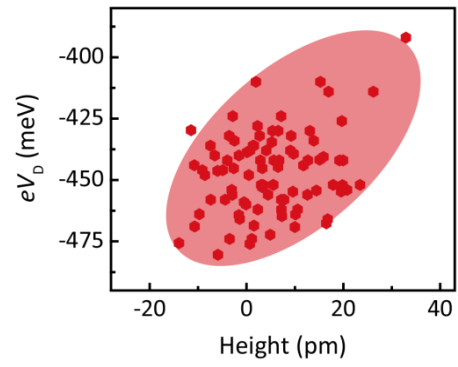

**FIG. S1. Statistical correlation between $eV_D$ and height.** The data point are extracted from Figs. 2(b), 2(d) and 2(f), appearing positively correlated more clearly in statistics. The shaded ellipse is the guide to the eyes of the statistical positive correlations between $eV_D$ and height.

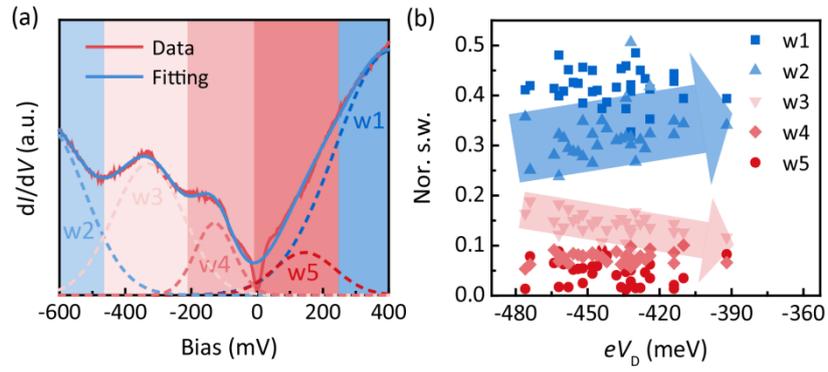

**FIG. S2. Full information of the multi-Gaussian fit.** (a) Exemplifying multi-Gaussian fit (solid blue curve) to the tunneling spectrum (solid red curve). The 'phonon' gap is ignored in the fit procedure. The involved five Gaussian components with integrated spectral weights w1–w5 are shown in dashed curves. Note that w2 (w3) is equal to $w_{LB}$ ($w_{UB}$) as defined in main text. (c) Normalized spectral weights (s.w.) w1–w5 extracted from the spectra in (a) plotted vs. $eV_D$. Clearly, while w2 and w3 show positive and negative correlations with $eV_D$, respectively, w1, w4, and w5 behave $eV_D$-independent statistically, highlighting the truly existent spectral-weight transfer between w2 and w3.

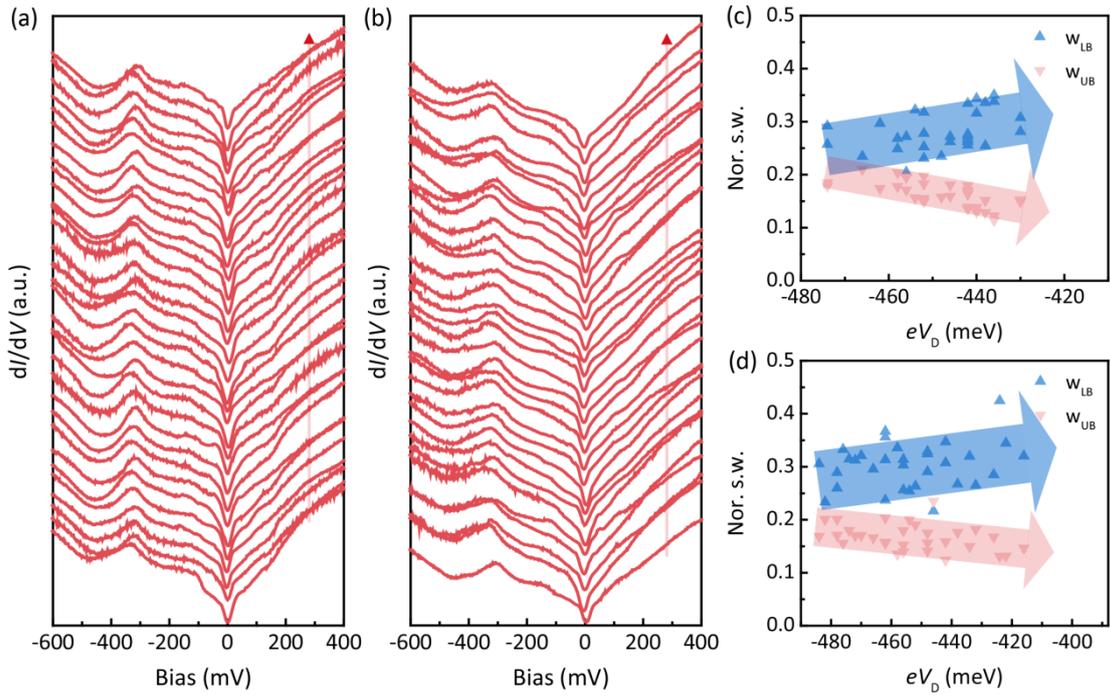

**FIG. S3. Reproducibility of the spectral-weight transfer in BLG.** (a,b) Tunneling spectra (vertically offset for clarity) taken along the arrows in Figs. 2(a) and 2(e), respectively. (c,d) Normalized spectral weights $w_{LB}$ and $w_{UB}$ extracted from the spectra in (a,b), respectively, plotted vs. $eV_D$.

14